\documentclass[11pt]{article}
\usepackage[english]{babel}
\usepackage[T2A]{fontenc}
\usepackage{amsmath,amsfonts}
\usepackage{amssymb}
\usepackage{amsxtra}
\usepackage{textcomp,color}
\usepackage{graphics,graphicx}
\usepackage{hyperref}
\usepackage{overpic}
\usepackage{setspace}
\usepackage{subcaption}
\usepackage[style=gost-numeric,bibencoding=auto,backend=biber,sorting=none,language=auto,autolang=other,doi=false,url=false,isbn=false]{biblatex}
\usepackage{caption}
\title{Dynamics of a homogeneous and isotropic space in pure cubic $f(R)$ gravity}

\author{Polina Petriakova \\ National Research Nuclear University MEPhI, \\ 115409, Kashirskoe shosse 31, Moscow, Russia \\ petriakovapolina@gmail.com}

\begin{document}
\maketitle
\begin{abstract}
We consider a possible ways of the dynamics of a homogeneous and isotropic space described by the Friedmann{-}Lemaitre{-}Robertson{-}Walker metric in the framework of cubic $f(R)$ gravity in the absence of matter. This article points an method for limiting the parameters of extended gravity models. We propose and develop a method for f(R) gravity models based on the dynamics of metrics for various model parameters in the simplest example. The influence of parameters and initial conditions on further dynamics are discussed. The parameters can be limited by 1) slow growth of space, 2) instability, 3) divergence with the inflationary scenario.
\end{abstract}

\section{Introduction}\label{Intro}
Despite the success of experimental tests of the Theory of General Relativity (GR) with excellent accuracy [1-3] 
the study of various modifications of theory of gravity continues still up to date. Historically, the first attempts at modifying GR were aimed at unification of gravity with other interactions by adding higher dimensions [4,5]. 
Modern interest in modified gravity has increased with the emergence of a large set of observational cosmology data [6]. 
The rapid development of experimental cosmology has cast doubt on the Big Bang theory. The standard cosmology of the Big Bang was well described in the framework of GR. In the case of a homogeneous and isotropic space the Einstein's equations lead to the Friedmann solutions [7] 
which describe the stages of dominance of radiation and matter. 

But modern observational data indicate existence stages of accelerated expansion of the Universe. The first is the inflation hypothesis, which is required not only to solve flatness and horizon problems, but also to explain the nearly flat temperature anisotropy spectrum observed in the cosmic microwave background [8]. 
The second is the modern accelerated expansion stage [9,10]. These two stages of the accelerated expansion of the Universe can not be explained in terms of standard matter with the known equation of state in the framework of GR. But these phenomena can be explained in the framework of modified gravity.

One of the simplest approach of modified gravity is $f(R)$ gravity. This class of theories is widely used in modern research [11-14] 
and in some cases successfully solves specific problems and fits the observational cosmology  data [15-18]. 

The first one and most successful formulation belonging to the $f(R)$ class of theories was the Starobinsky model [19] 
containing only one free parameter. In this model, the addition of the $R^2${--}term was made for elimination of cosmological singularity and led to the inflationary stage. The Starobinsky's inflationary model is a particular solution of the class of theories of gravity with higher derivatives which are devoid of ghost degrees of freedom, perturbatively unitary and finite at the quantum level [20]. 
This model has a "graceful exit" from inflation and provides a mechanism for subsequent creation and final thermalization of the standard matter. 
However, adding a cubic term may provide better agreement with inflationary data, as recently shown in [22]. 

In this paper, it is proposed to study the possible ways of evolution of a homogeneous and isotropic space in the framework of cubic $f(R)$ gravity. Considering only the gravitational component of the evolution we discuss the influence of parameters and initial conditions on further dynamics.

\section{Basic equations}\label{BasicEq}
Let us consider the theory with the following action \footnote{We use the rationalized Planck units $ \hbar = c = k_B = 8 \pi G = 1$, hence further $m_{\text{Pl}}=1$.}
\begin{equation}\label{act0}
S[g_{\mu \nu}]=\frac{m_{\text{Pl}}^{2}}{2} \int d^4 x \sqrt{|g|}\, f(R) \, .
\end{equation}
By varying this action with respect to the metric and  using conventions for the curvature tensor as $R_{\mu\nu \alpha}^{\beta}=\partial_{\alpha}\Gamma_{\mu \nu}^{\beta}-\partial_{\nu}\Gamma_{\mu \alpha}^{\beta}+\Gamma_{\sigma \alpha}^{\beta}\Gamma_{\nu\mu}^{\sigma}-\Gamma_{\sigma \nu}^{\beta}\Gamma_{\mu \alpha}^{\sigma}$ and the Ricci tensor is $R_{\mu \nu}=R^{\alpha}_{\mu \alpha \nu }$, we~get the equation of motion for the $f(R)$ gravity theory
\begin{equation}\label{eqf(R)}
f_{R}(R) R_{\mu \nu} - \frac{1}{2} \, f(R) g_{\mu \nu} + \Bigl[ \nabla_{\mu} \nabla_{\nu} - g_{\mu \nu} \Box \Bigr]f _{R} (R)=0\, ,
\end{equation}
here $\Box \equiv g^{\mu \nu} \nabla_{\mu} \nabla_{\nu}$ and $f_R (R) \equiv  d f(R)/ d R $.

Taking into account the choice of the metric of a homogeneous and isotropic spaces
\begin{align}\label{ds3dimPos}
    ds^{2}_{+} &= dt^2 - \text{e}^{2 \alpha(t)}\Bigl(dx^2 + \sin^2{x} \,  dy^2 + \sin^2{x} \, \sin^2{y} \,  dz^2\Bigr)\, , \\
    \label{ds3dimFlat}
    ds^{2}_{0} &= dt^2 - \text{e}^{2 \alpha(t)}\Bigl(dx^2 + dy^2 + dz^2\Bigr)\, , \\
    \label{ds3dimNeg}
    ds^{2}_{-} &= dt^2 - \text{e}^{2 \alpha(t)}\Bigl(dx^2 + \sinh^2{x} \,  dy^2 + \sinh^2{x} \, \sin^2{y} \,  dz^2\Bigr)\, ,
\end{align}
which corresponds to spaces with positive \eqref{ds3dimPos}, zero \eqref{ds3dimFlat} and negative \eqref{ds3dimNeg} curvature, we obtain from \eqref{eqf(R)} the nontrivial equations
\begin{align}\label{eqm_tt}
&6\dot{\alpha}\dot{R}f_{RR} (R)-6 \Bigl(\ddot{\alpha}+\dot{\alpha}^2 \Bigr)f_R (R) +f(R) = 0 \, ,\\
\label{eqm_xx}
&2\dot{R}^2 f_{RRR}(R)+2\Bigl(\ddot{R}+2\dot{\alpha}\dot{R}\Bigr)f_{RR}(R) -\Bigl(2\ddot{\alpha}+6\dot{\alpha}^2+4 \gamma \text{e}^{-2\alpha(t)} \Bigr)f_R (R) + f(R)=0 \, ,
\end{align}
where the Ricci scalar for used metrics is
\begin{equation}\label{R}
    R(t) = 12\dot{\alpha}^2(t) + 6\ddot{\alpha}(t) + \gamma 6\text{e}^{-2\alpha(t)}
\end{equation}
with $\gamma = +1$ for \eqref{ds3dimPos}, $\gamma = 0$ for \eqref{ds3dimFlat} and $\gamma = -1$ for \eqref{ds3dimNeg}. 
 
In order to find a solution, a system will determined consisting of \eqref{eqm_xx} and the definition the Ricci scalar \eqref{R}. Equation \eqref{eqm_tt} will be used as a constraint on the initial condition. Let us choose the initial conditions for the unknown functions $\alpha(t)$ and $R(t)$ of the system of equations \eqref{eqm_xx} and \eqref{R} as
\begin{equation}\label{const_cond}
\alpha(0) = \alpha_0 \, , \quad \dot{\alpha}(0) = \alpha_1 \, , 
\quad \dot{R}(0) = R_1 \, 
\end{equation}
and the initial value of the curvature $ R(0) = R_0$ will be found by solving the equation~\eqref{eqm_tt}.

The most difficult question is the choice of the initial conditions. In our formulation of the problem the conditions \eqref{const_cond} are free parameters that we set "by hand". 
In addition to the problem of choosing the initial conditions, some models contain the possibility of several asymptotic values of the Ricci scalar for a specific form of the $f(R)$ function. We are going to consider this situation in the simplest case when the chosen function has the form
\begin{equation}\label{f(R)3}
    f(R)=a_3 R^3 + a_2R^2 + R \, .
\end{equation}
In this case, the equation \eqref{eqf(R)} at the constant curvature $R_c = const$ leads to equation
\begin{equation}\label{eqf(Rc)}
f ' _{R} ( R_{c} ) R_{c} - 2 \,f( R_{c} ) =0 \, ,
\end{equation}
which defines the asymptotic values of the Ricci scalar. For our simplest case when more than one asymptotic value is possible we get from \eqref{eqf(Rc)}
\begin{equation}\label{asympt}
    R_c = 0  \, , \,\, R_c = \frac{1}{\sqrt{\,a_3\,}} \, , \,\, \text{and}  \,\, R_c = - \frac{1}{\sqrt{\,a_3\,}} \, .
\end{equation}

So what conditions lead to the realization of a specific asymptotics?

\section{Analysis in the Einstein frame}\label{EPicture}

In this section, we are going to verify the effect of model parameter values and initial conditions on the dynamics of spaces. Action \eqref{act0} can be reduced to scalar-tensor theory by introducing an auxiliary scalar field $\chi$ as a result of the Legendre transformation: 
\begin{equation}\label{ST}
    S_{ST} = \frac{1}{2} \int  d^{4} x \sqrt{-g}\left[f(\chi) + f'(\chi) \left(R-\chi\right) \right]\, .
\end{equation}
After conformal transformation ${g}_{\mu \nu}=|f'(\chi)|^{-1} \, \hat{g}_{\mu \nu}$
we get this action in the Einstein frame
\begin{equation}\label{ST1}
S_{E}=\frac{1}{2}\int d^4 x \sqrt{-\hat{g}}\left[ \hat{R}+ \hat{g}^{\mu\nu}\partial_{\mu}\, \psi \partial_{\nu}\, \psi - 2V(\psi)\right] \, ,
\end{equation}
where new variables were introduced as follows
\begin{equation}\label{new_values}
\psi=\sqrt{\frac{3}{2}}\ln{f'(\chi)} \, ; \quad  V(\psi) = \left. \frac{\Bigl(\chi f'(\chi)-f(\chi)\Bigr)}{2 \bigl(f'(\chi)\bigr)^2}\, \,   \right|_{\chi=\chi(\psi)} \, 
\end{equation}
To avoid the antigravity regime, we come to the condition $f'(\chi)>0$ [24]. 
One of the classical equation of the action \eqref{ST} is $f''(\chi)\bigl(R-\chi\bigl)=0$ and then $R=\chi$ if $f''(\chi) \neq 0$. Using the form \eqref{new_values} of the potential $V(\psi)$  we can find a condition for a local extremum of an auxiliary scalar field
\begin{equation}\label{dV_chi}
  \frac{d V(\psi)}{d \chi} = - \left. \frac{f''(\chi)\Bigl(\chi f'(\chi)- 2 f(\chi)\Bigr)}{2\bigl(f'(\chi)\bigr)^3}\right|_{\chi=\chi(\psi)} \xrightarrow[]{\, f''(\chi) \neq 0 \,} \chi f'(\chi)- 2 f(\chi)=0.
\end{equation}
 This condition \eqref{dV_chi} is exactly the condition \eqref{eqf(Rc)} obtained before, which leads to a de Sitter space endowed with constant curvature.
 
The existence of a stable minimum for a scalar field follows from
\begin{equation}\label{ddV_chi}
  \frac{d^2 V(\psi)}{d \chi^2} = \left. \frac{f''(\chi) \Bigl(- \chi f''(\chi) + f'(\chi)\Bigr)}{2\bigl(f'(\chi)\bigr)^3}\right|_{\chi=\chi(\psi)} > 0 \, \rightarrow \, - \chi + \frac{f'(\chi)}{f''(\chi)} > 0 \, .
\end{equation}
From the obtained condition  \eqref{ddV_chi} we get for \eqref{f(R)3}
\begin{equation}\label{Rc_fR3_a0}
    \begin{matrix}
    - R_c + \cfrac{3a_3R_{c}^{2}+2a_2R_c+1}{6a_3R_c+2a_2} &=
    & \left\{
    \begin{matrix}
    \cfrac{1}{2a_2}, & \mbox{for} & R_c = 0\, , \\
    -\cfrac{1}{a_2 \pm 3\sqrt{a_3}}\, , & \mbox{for} &R_c = \pm \cfrac{1}{\sqrt{a_3}} .
    \end{matrix} \right.
    \end{matrix}
\end{equation}

\begin{figure}[!h]
\centering
 \begin{subfigure}[t]{0.35\linewidth}
 \includegraphics[width=\textwidth]{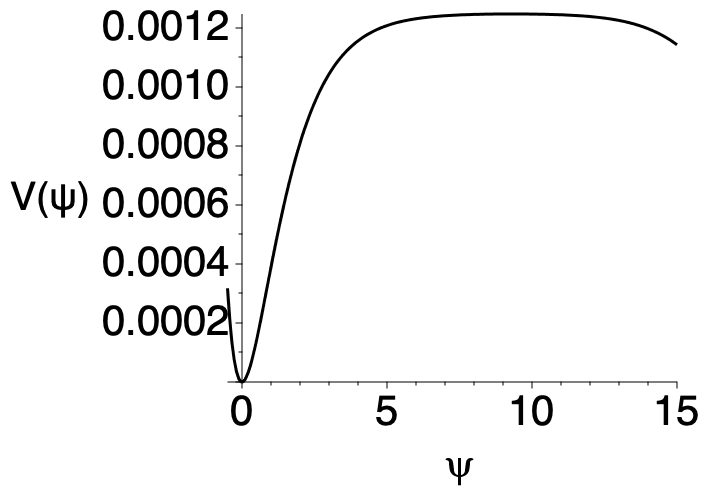}
 \subcaption{}
 \includegraphics[width=\textwidth]{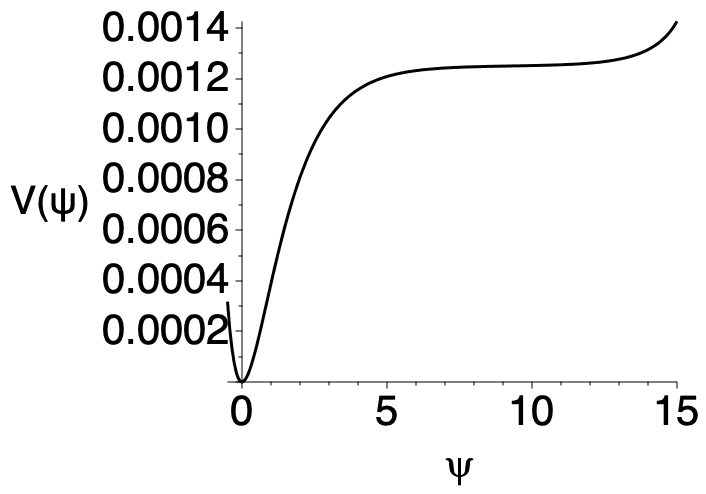}
 \subcaption{}
 \end{subfigure}
 \begin{subfigure}[t]{0.35\linewidth}
 \includegraphics[width=\textwidth]{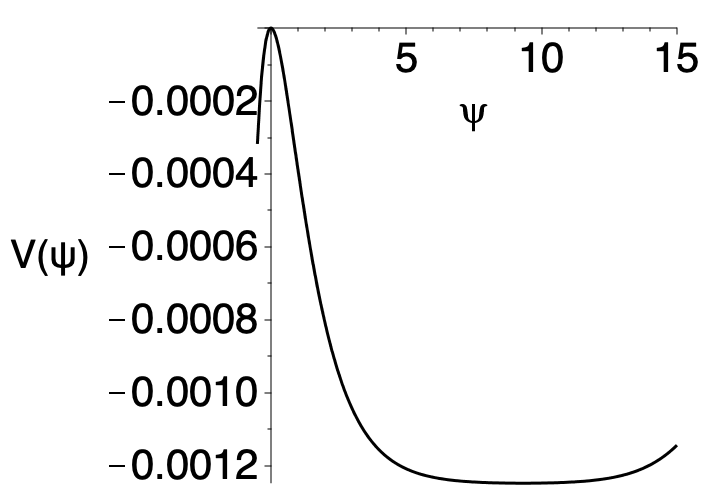}
 \subcaption{}
 \includegraphics[width=\textwidth]{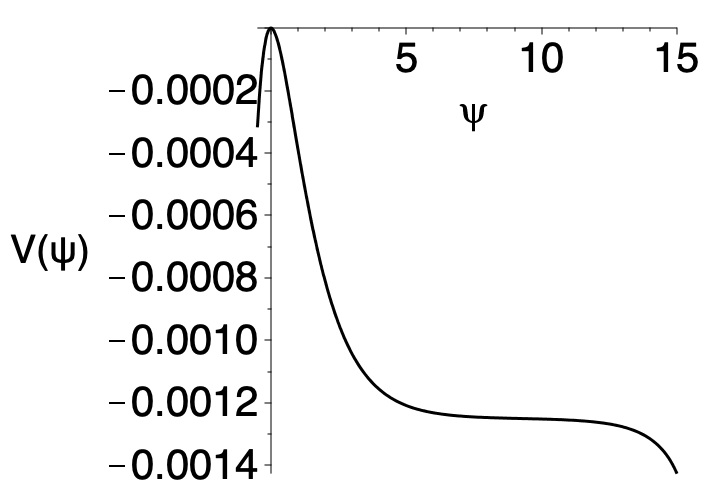}
 \subcaption{}
 \end{subfigure}
\caption{The potential $V(\psi)$ \eqref{new_values} with the chosen values of parameters of the function \eqref{f(R)3} as $a_3 = 0.01$, $a_2 = 100$ (a); $a_3 = - 0.01$, $a_2 = 100$ (b); $a_3 = 0.01$, $a_2 = -100$ (c) and $a_3 = - 0.01$, $a_2 = -100$ (d).}
\label{Fig:Potentials}
\end{figure}

One of the asymptotics corresponds to the maximum of the potential, while the other corresponds to the minimum depending on the values of the parameters of the function \eqref{f(R)3}. An illustration of the potential depending on the signs of the parameters of the function \eqref{f(R)3} is shown in Figure~\ref{Fig:Potentials}. We should remind that the introduced value of $\psi$ is related to the auxiliary scalar field $\chi$ by formula \eqref{new_values} and $\chi = R$.

 An unstable position is possible at large $\psi$ values in the (a) and (d) pictures in Figure~\ref{Fig:Potentials} where both coefficients $a_3$ and $a_2$ have the same sign. The initial conditions in situation at the (a) picture can play an important role and lead to an unstable solution. The chosen initial conditions \eqref{const_cond} and values of the coefficients must lead to a value $R_0< R_c = 1 /\, \sqrt{a_3}$ \, to ensure the stability of the solution. While instability in the (d) picture arises regardless of the choice of initial conditions. The dynamics in the case of the other two forms of potential which corresponds the (b) and (c) pictures in Figure~\ref{Fig:Potentials} is quite predictable. Regardless of the initial value of the curvature, the solution will tend to a stable minimum and reach it.
 
 Thus, we have revealed the influence of the values of the parameters of the $f(R)$ function \eqref{f(R)3} determining the implementation of the asymptotics \eqref{asympt}. The only exception in this chosen model \eqref{f(R)3} is the case of the (a) picture where the initial conditions can lead to unstable solutions.

In the next section we confirm the obtained statements by numerical calculations.

\section{Numerical results}
Let us illustrate what was discussed above with the example of a numerical solution for a flat space \eqref{ds3dimFlat}. As mentioned in Section \ref{BasicEq} we are going to solve a system of equations \eqref{eqm_xx} and \eqref{R} with initial conditions \eqref{const_cond} chosen near the sub{-}Planck scale
\begin{equation}\label{c1}
\alpha_0 \sim -\ln{H_{\text{sub-Planck}}} \,\, , \quad  \alpha_1\, \sim H_{\text{sub-Planck}} \,\, , \quad R_1 = 0 \, ,
\end{equation}
here $H_{\text{sub-Planck}}\, \lesssim \, m_{\text{Pl}}$ and the initial value of the curvature $ R_0$ will found from solution of equation \eqref{eqm_tt}.

\begin{figure}[!h]
\centering
\includegraphics[width=0.32\textwidth]{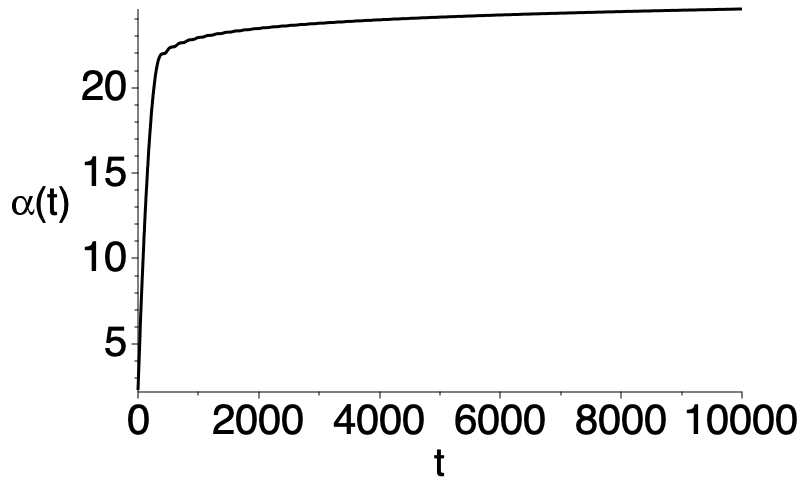}
\includegraphics[width=0.32\textwidth]{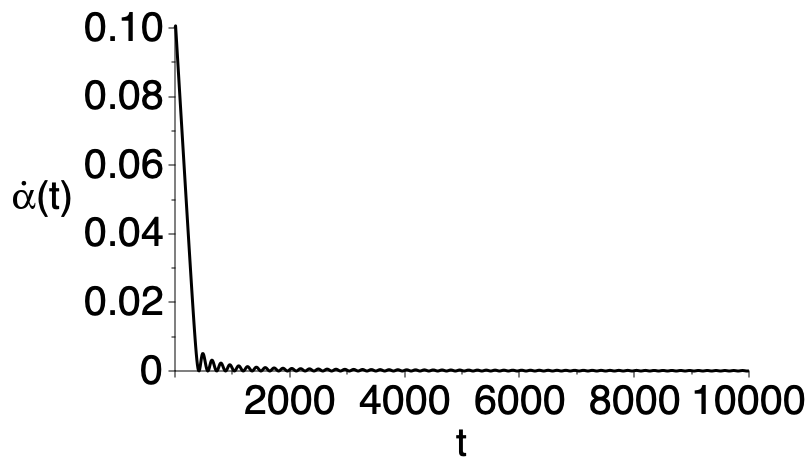}
\includegraphics[width=0.32\textwidth]{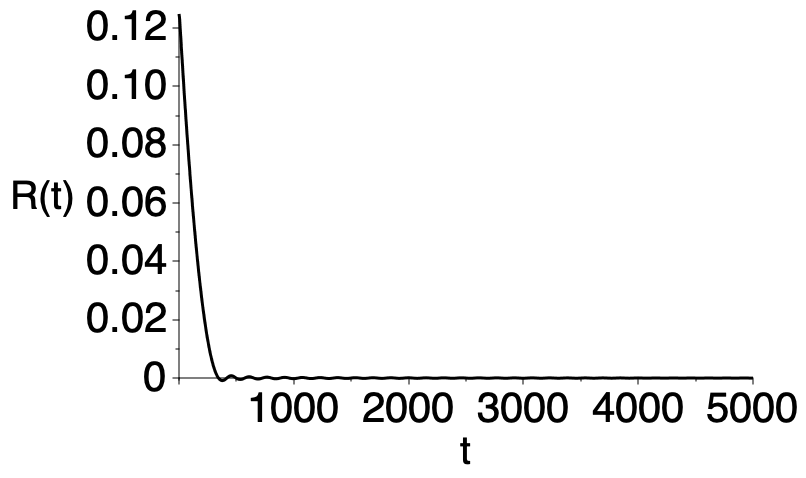}
\caption{ The solution of the system of equations \eqref{eqm_xx} and \eqref{R} with parameters $a_3=0.01$, $a_2=100$  with the initial conditions $ \alpha_0=2.3 $, $ \alpha_1 = 0.1 $, $R_0=0.125$ and $R_1=0$. The asymptotic behavior is $ R_c=0 $, $ H \equiv \dot{\alpha}(t_{\infty})=0 $. The parameter values $a_3$ and $a_2$ correspond to the picture (a) in Figure \ref{Fig:Potentials}.}
\label{Fig:FlatDyn}
\end{figure}

\begin{figure}[!h]
\centering
\includegraphics[width=0.32\textwidth]{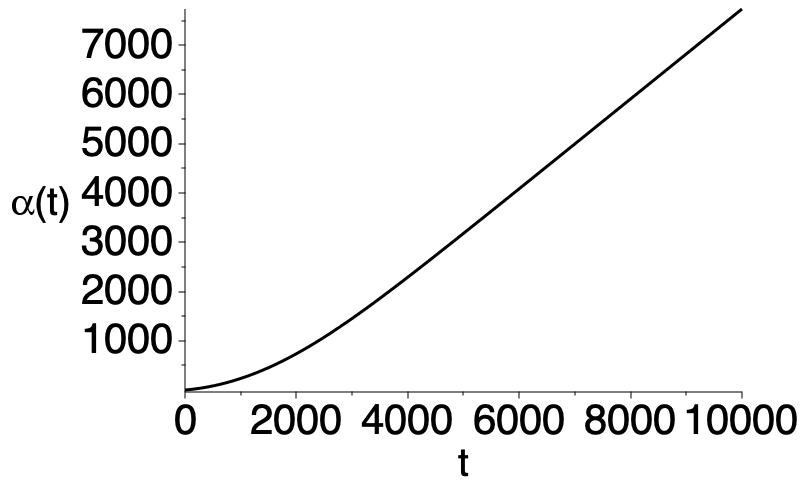}
\includegraphics[width=0.32\textwidth]{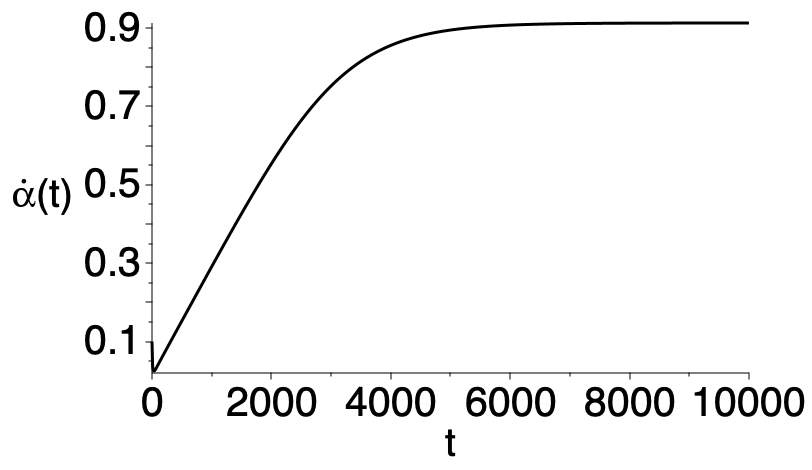}
\includegraphics[width=0.32\textwidth]{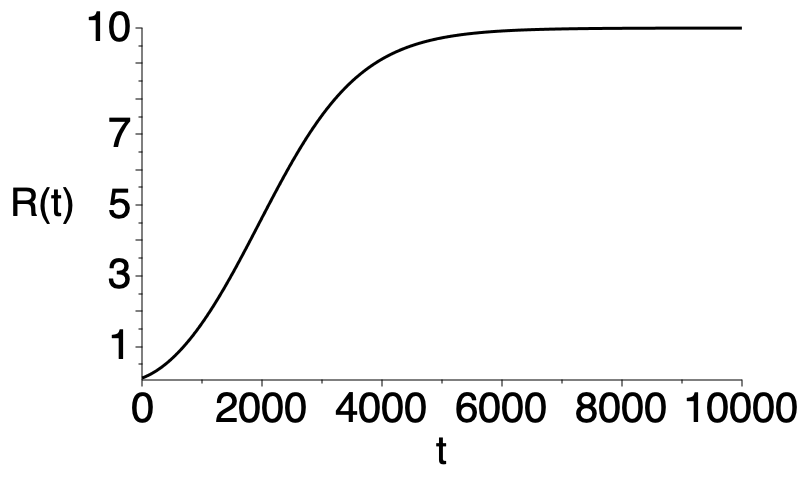}
\caption{ The solution of the system of equations \eqref{eqm_xx} and \eqref{R} with parameters $a_3=0.01$, $a_2=-100$ and initial conditions $ \alpha_0=2.3 $, $ \alpha_1 = 0.1 $, $R_0=0.005$ and $R_1=0$. The asymptotic behavior is $ R_c=10 $, $ H \equiv \dot{\alpha}(t_{\infty})=0.91 $. The parameter values $a_3$ and $a_2$ correspond to the (c) picture in Figure~\ref{Fig:Potentials}.} 
\label{Fig:FlatDyn2}
\end{figure}

The results of numerical solution for this Cauchy problem in the rationalized Planck units are presented in Figures \ref{Fig:FlatDyn} and \ref{Fig:FlatDyn2}. We see in Figure~\ref{Fig:FlatDyn} a direct correspondence to the case in the picture (a) of Figure~\ref{Fig:Potentials} and  Figure~\ref{Fig:FlatDyn2} corresponds to the (c) case in Figure~\ref{Fig:Potentials}.

In Figure \ref{Fig:FlatDyn} the numerical solution covers several stages of the evolution of the Universe. The first is the inflationary stage where space grows exponentially and the curvature of space decreases. Then, having reached zero, the curvature begins to oscillate around zero and the Hubble parameter, i.e. $\dot{\alpha}(t)$, at this stage reaches asymptotically zero and the size of the space $\alpha(t)$ tends to a constant. After the damping of the oscillations of the corresponding quantities a transitional regime begins to the GR. But the space does not expand enough to describe the visible part of the Universe. The value must be $\alpha(t_{\infty}) > 140$ where $\text{e}^{140}m^{-1}_{\text{Pl}}$ corresponds to the horizon scale $\sim 10^{28}$ cm at present time. In addition to the final size of space, we see that the exponential expansion of space ends earlier than the corresponding duration of the inflationary stage. The exponential expansion of space must continue $ \delta t \sim 10^7$ in the used units\footnote{The rationalized Planck units $ \hbar = c = k_B = 8 \pi G = 1$, hence $m_{\text{Pl}}=1$.} and during this time the function $\alpha(t)$ must change by the value $ \Delta \alpha(\delta t) = N_{\text{e}} \approx 60$.

 \newpage
In Figure \ref{Fig:FlatDyn2} space begins its evolution in the early stage in a similar way, but then enters a different asymptotic behavior which is dictated by the signs of the coefficients and does not correspond to the observable Universe. As a result, we get an infinitely exponentially growing space endowed with significant curvature.

Specific values of the parameters $a_3$ and $a_2$ of the $f(R)$ function determine the duration of each stage of evolution and the amplitudes of oscillations that have cosmological significance. In addition to the revealed dependence of the realization of asymptotics on the parameters of the $f(R)$ function, the rate of dynamics of space is also important issue. The~previous reasoning and conclusions in Section \ref{EPicture} did not depend on the specific choice of the metric. Let us find a numerical solution for three possible configurations of a homogeneous and isotropic space \eqref{ds3dimPos}, \eqref{ds3dimFlat} and \eqref{ds3dimNeg}. For more realistic results describing the observable Universe we will choose the parameters following the constraints from work [22] 
as $a_3=-10^3$ and $a_2 = 10^9$ based on analysis of the inflationary scenario. 

\begin{figure}[!h]
\centering
\includegraphics[width=0.32\textwidth]{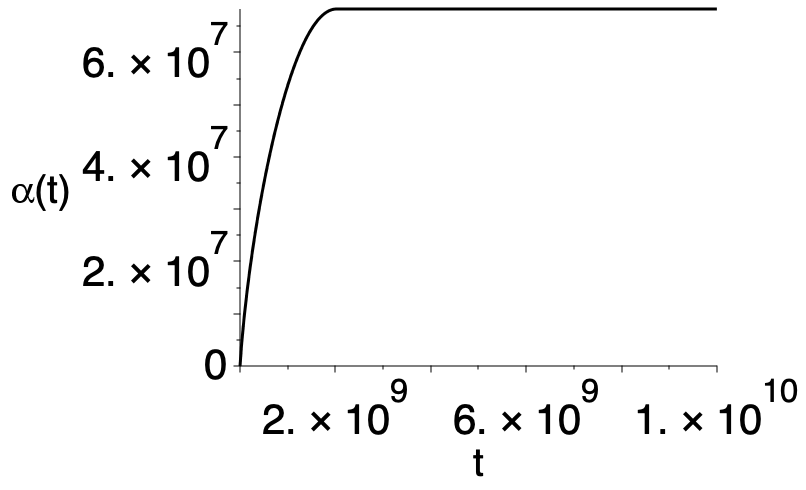}
\includegraphics[width=0.32\textwidth]{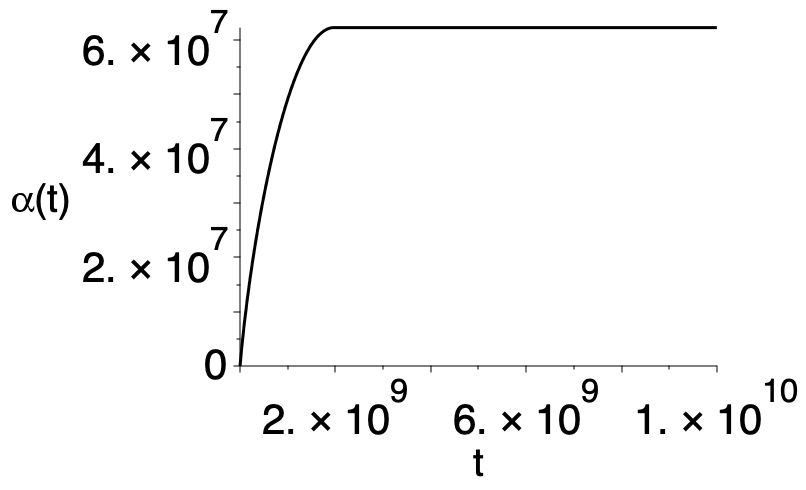} 
\includegraphics[width=0.32\textwidth]{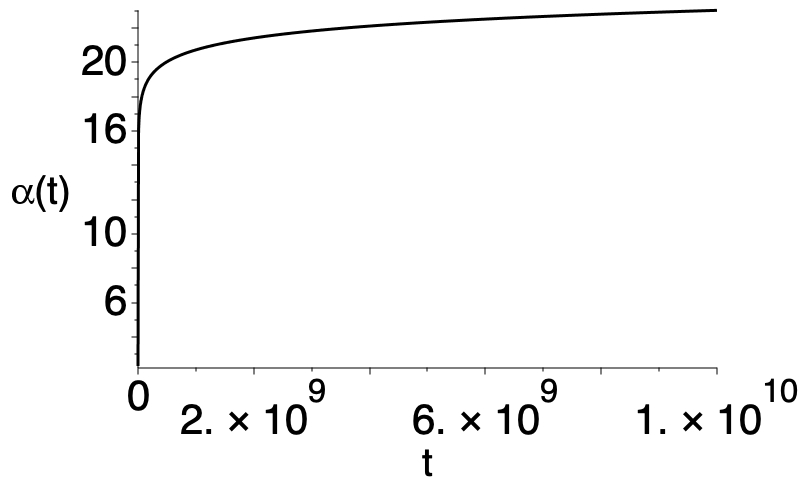} 
\caption{ The solution of the system of equations \eqref{eqm_xx} and \eqref{R} with parameters $a_3=-10^3$, $a_2=10^9$ and initial conditions $ \alpha_0=2.3 $, $ \alpha_1 = 0.1 $ and $R_1=0$ for the metric \eqref{ds3dimPos} (first), \eqref{ds3dimFlat} (second) and \eqref{ds3dimNeg} (third), $R_0$ for each case defined from solution of equation \eqref{eqm_tt}.}
\label{Fig:AllDyn}
\end{figure}

The results of the dynamics of spaces of different curvature with the same choice of parameters and initial conditions \eqref{c1} are shown in Figure~\ref{Fig:AllDyn}.

\newpage 
The obtained results indicate a different rate of dynamics of spaces. We see that the behavior of a metric with positive curvature \eqref{ds3dimPos} and flat \eqref{ds3dimFlat} leads to a value of the same order of magnitude in the asymptotics. While the dynamics of space of negative curvature \eqref{ds3dimNeg} leads to a size less than the visible part of the Universe.

\begin{figure}[!h]
\centering
\includegraphics[width=0.32\textwidth]{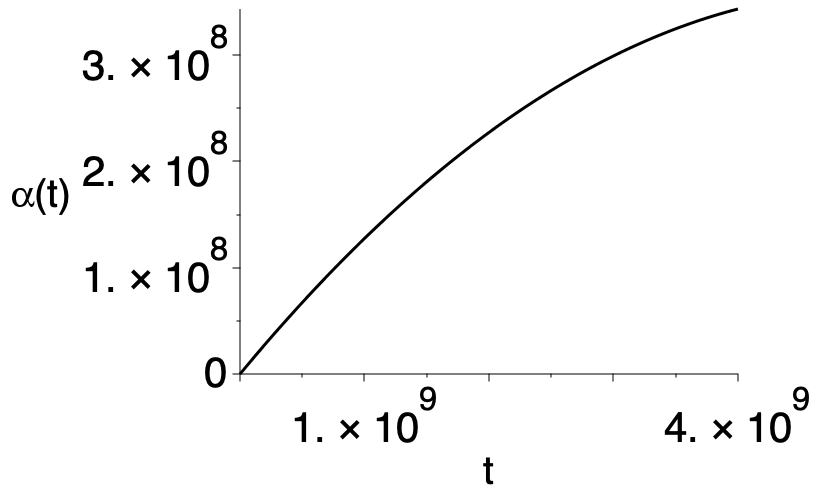}
\includegraphics[width=0.32\textwidth]{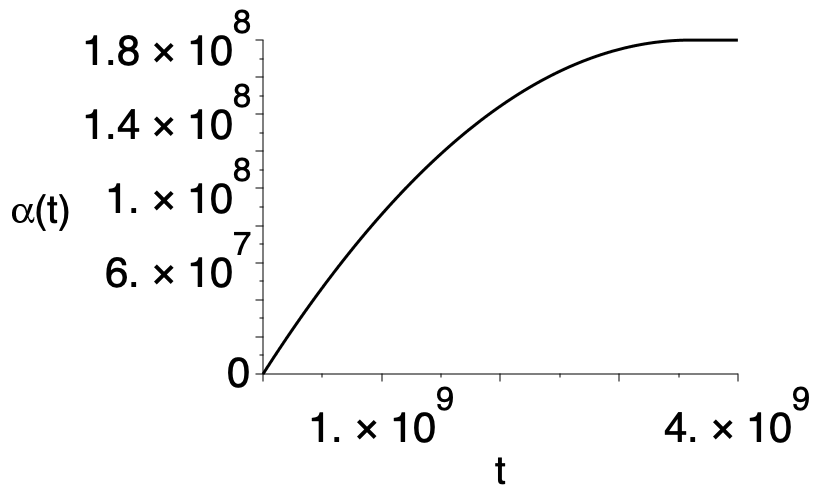}
\includegraphics[width=0.32\textwidth]{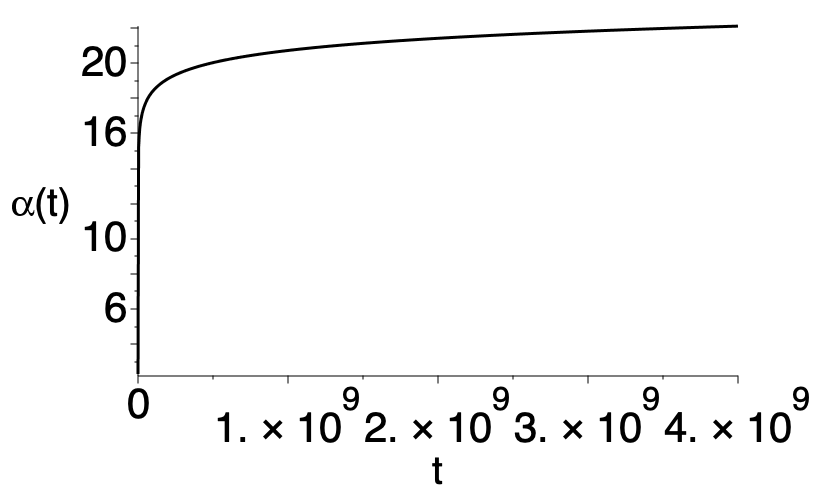}
\caption{ The solution of the system of equations \eqref{eqm_xx} and \eqref{R} with parameter $a_2 \simeq 10^9$ with the initial conditions $ \alpha_0=2.3 $, $ \alpha_1 = 0.1 $ and $R_1=0$ for the metric \eqref{ds3dimPos} at the left side, \eqref{ds3dimFlat} in the middle and \eqref{ds3dimNeg} at the right side, $R_0$ for each case defined from solution of equation \eqref{eqm_tt}.}
\label{Fig:AllDynSt}
\end{figure}

A similar difference in the rate of expansion of space takes place in the framework of the Starobinsky model [19] 
\begin{equation}
    f(R)= \frac{1}{6m^2} R^2 + R \, , \quad m \sim 1.5\cdot 10^{-5} \, m_{\text{Pl}}\, \biggl(\cfrac{50}{N_\text{e}}\biggr), \quad N_{\text{e}} = 55 \div 60 \, .
\end{equation}
The results of such dynamics of the spaces \eqref{ds3dimPos}, \eqref{ds3dimFlat} and \eqref{ds3dimNeg} are presented in Figure \ref{Fig:AllDynSt}.

\section{Discussion}
Considering only the gravitational dynamics of a homogeneous and isotropic space, we come to restrictions on $f(R)$ function. 
The asymptotics is strictly determined by the signs and values of the coefficients. Nevertheless, the choice of the initial conditions can affect to the stability of the solution. In addition to the correct asymptotic value of the curvature, it is also worth considering the possible size of space, no less than the size of the visible part of the Universe and the values of cosmological parameters available from the observational data [6]. 
A detailed analysis of cubic gravity in the inflationary scenario was recently performed in [22]. 
It was shown that such an extension of Starobinsky’s model allows to get better agreement with experimental data. However, considering the purely gravitational dynamics of an isotropic and homogeneous space [23] 
the acceptable range of values of the coefficients of $f(R)$ function has intersection with the range of values from [22] 
less than each of them provide separately.

Besides, at Planck energy scales fluctuations can lead to the formation of asymmetric spaces. The description of this process can be done only by the theory of quantum gravity that has not yet been developed. Spaces of different curvatures have different rates of expansion as a result of our numerical solutions of the classical equations of motion. Then, can our Universe be homogeneous and isotropic space outside the visible part? This issue will be our further work.

\section*{Acknowledgments}
I would like to thank Prof. Sergey G. Rubin for the usefull discussions and interest to the work. The work was supported by the Ministry of Science and Higher Education of the Russian Federation, Project "Fundamental properties of elementary particles and cosmology" No 0723-2020-0041. This work was as part of the master course "Cosmoparticle physics" in National Research Nuclear University MEPhI.

\end{document}